\documentclass[11pt]{article}
\usepackage[T1]{fontenc}
\usepackage[utf8]{inputenc}
\usepackage{mathptmx}
\usepackage[scaled=0.92]{helvet}
\usepackage{courier}
\usepackage[margin=1in]{geometry}
\usepackage{amsmath,amssymb,mathtools,bm}
\usepackage{booktabs,tabularx,array,multirow,longtable,threeparttable}
\usepackage{graphicx,float}
\usepackage{microtype}
\usepackage{url}
\usepackage{authblk}
\usepackage[hidelinks]{hyperref}

\title{Discrete proportional-hazards modelling of lithium-ion battery end of life under interval inspection}
\author[1]{Simhaa T. T.\thanks{Corresponding author: simhaa2310510@ssn.edu.in}}
\author[2]{Sudheesh Kumar Kattumannil}
\affil[1]{Department of Electrical and Electronics Engineering, Sri Sivasubramaniya Nadar College of Engineering, Kalavakkam, Chennai 603110, Tamil Nadu, India}
\affil[2]{Applied Statistics Unit, Indian Statistical Institute, Chennai Centre, Chennai 600029, Tamil Nadu, India}
\date{}
\begin{document}
\maketitle
\begin{abstract}
Capacity-defined battery end of life is observed at discrete inspections, often with incomplete follow-up. We study a penalised complementary-log-log proportional-hazards model for such records and distinguish interval-hazard estimation from fixed-horizon prediction at a prespecified landmark. Six ageing sources provide 493 cells, with 303 harmonised endpoint events and 190 censored trajectories. At the selected ridge penalty, a one-standard-deviation increase in lagged state of health was associated with a hazard ratio of 0.50 (95\% whole-cell bootstrap interval 0.47--0.52). In an independent 42-cell MATR evaluation, separate leakage-free landmark cloglog regressions gave Brier scores from 0.0366 to 0.0967; a random forest had a lower mean Brier score over nine matched settings (0.0589 versus 0.0732). In a grouped proportional-hazards simulation, cloglog recovered the generating coefficient with less bias than logit or probit despite similar probability errors. Blind transfer to NMC overpredicted interval risk (0.074 versus 0.026 observed). Absolute-risk calibration should therefore be checked on the target population before the model informs replacement decisions.
\end{abstract}

\noindent\textbf{Keywords:} discrete-time proportional hazards, battery end of life, interval inspection, right-censoring, landmark prediction, calibration, reliability modelling

\section{Introduction}\label{sec:intro}
Condition-based maintenance decisions require more than a single predicted remaining life. For a cell observed at cycle $L$, the pertinent quantity may be the probability of reaching a capacity threshold before cycle $H$, conditional on the information already available. Such probabilities inform service intervals, screening for reuse and comparisons between replacement rules. Throughout this article, end of life (EOL) denotes a specified capacity-retention endpoint, not thermal runaway or other acute safety failure. The analysis concerns individual cells rather than an assembled battery pack.

A second issue is how failure time is recorded. Battery capacity is usually measured at cycles or diagnostic inspections rather than continuously. When testing stops before EOL, the available trajectory is right-censored. A grouped proportional-hazards (PH) model assigns a conditional event probability to each inspected interval and connects that probability to the corresponding cumulative-hazard increment \cite{PrenticeGloeckler1978,Allison1982,TutzSchmid2016}. It offers a direct alternative to treating the first recorded crossing as an exact continuous failure time. The complementary-log-log (cloglog) parameterisation also preserves a PH coefficient when covariates remain fixed within an interval under an underlying continuous PH process.

The modelling choice differs from first-passage methods. Particle filtering, Wiener-process and related stochastic degradation models infer a path towards an EOL boundary and derive a remaining-life distribution from it \cite{Dalal2011,Han2023,Yang2024,ZhangSiHuLei2018}. Hazard regression works with the conditional event probability instead of specifying that latent path. Both approaches belong to the Part O reliability literature, including condition-monitoring and maintenance analyses \cite{CarrWang2008,Bharat2018,Moniri2019,ZhangSiHuKong2015}. Neither formulation is automatically preferable: degradation-process predictions depend on drift, diffusion and measurement-error assumptions, while a PH interpretation relies on the suitability of the time-constant covariate coefficient. Shikdar and Laaksonen \cite{Shikdar2026} used multihorizon discrete-time hazard learning on 37 cells to estimate failure probability within a service window and to limit battery dispatch. The present study differs in using a proportional-hazards formulation whose coefficients are hazard ratios, a capacity-defined endpoint across six sources, and tests of transfer and recalibration.

Probability calibration is as important as discrimination when decisions use a risk threshold. A high receiver-operating-characteristic area (AUC) can coexist with systematic overprediction, and measurements made just before an endpoint can produce an overly favourable retrospective assessment \cite{Steyerberg2010,VanCalster2019}. We therefore distinguish two estimands: the interval event hazard in a person-period dataset and the probability of EOL at a future horizon predicted using only data observed at a landmark. The empirical benchmark uses a whole-cell train/test separation, while simulations isolate the effects of interval width, lead time and censoring. Source and chemistry holdouts assess the stability of the probability scale under population change.

The contributions are (i) an auditable, six-source person-period representation with a consistent capacity-endpoint rule; (ii) penalised grouped cloglog estimation and cell-bootstrap uncertainty for a lagged-health effect; (iii) a leakage-free landmark benchmark with comparisons of discrimination and proper probability scores; and (iv) link-function and recalibration studies that explain when an interval-hazard model retains a meaningful effect scale and when transferred probabilities require caution. The experiments are interpreted at their respective observation units rather than combined into a single performance claim.

\section{Battery data and endpoint definition}\label{sec:data}
\subsection{Six-source event-history corpus}
The harmonised corpus combines MATR with BatteryLife SNL, CALB, HUST, SDU and Tongji \cite{Severson2019,Tan2025BatteryLife}. Table~\ref{tab:sourceaudit} records the distinction between each source's published life labels and the harmonised endpoint. In total, 493 cells contribute 16,538 at-risk 25-cycle intervals after cycle 100, comprising 303 confirmed endpoints and 190 right-censored histories.
\begin{table}[tbp]\centering\small
\caption{Battery-source support. Native life labels are not interchangeable with confirmed harmonised capacity endpoints.}\label{tab:sourceaudit}
\begin{tabular}{lrrrr}\toprule
Source & Cells & Native labels & Events & Censored\\\midrule
MATR/Severson & 139 & 83 & 43 & 96\\
BatteryLife SNL & 61 & 52 & 34 & 27\\
BatteryLife CALB & 27 & 6 & 6 & 21\\
BatteryLife HUST & 77 & 77 & 77 & 0\\
BatteryLife SDU & 70 & 70 & 70 & 0\\
BatteryLife Tongji & 119 & 73 & 73 & 46\\\midrule
Total & 493 & 361 & 303 & 190\\\bottomrule
\end{tabular}
\end{table}
\begin{figure}[tbp]\centering
\includegraphics[width=\linewidth]{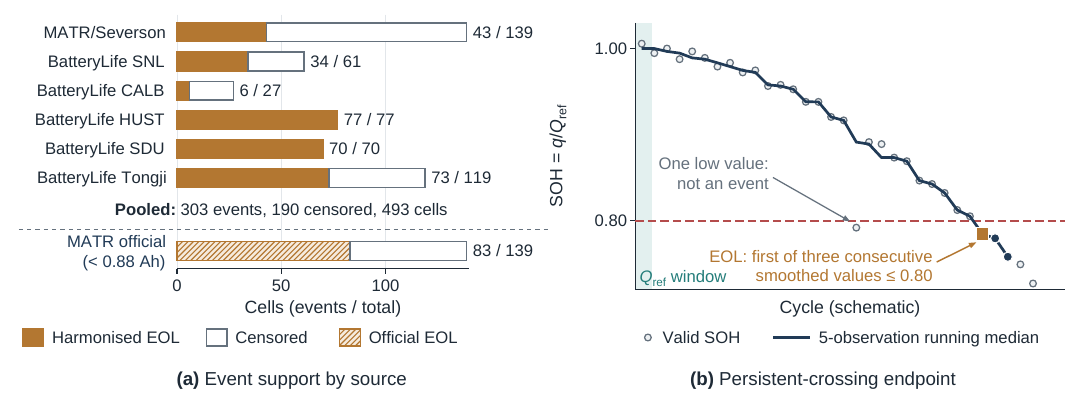}
\caption{Six-source event support and capacity-endpoint adjudication. Native life labels and harmonised events are shown separately; the capacity trace is an illustrative schematic, not a measured cell.}\label{fig:corpus_endpoint}
\end{figure}

For cell $i$, let $q_i(c)$ be measured discharge capacity and $Q_i^{\rm ref}$ the median of at least three valid measurements between cycles 2 and 10 (using the first five valid observations when necessary). The state of health is $s_i(c)=q_i(c)/Q_i^{\rm ref}$. With $\tilde s_i(c)$ denoting a centred five-observation median, the harmonised EOL is assigned at the first of three consecutive valid smoothed readings at or below $0.80$:
\begin{equation}\label{eq:pooled_eol}
T_i^{\rm harm}=\min\{c_k:\max_{r=0,1,2}\tilde s_i(c_{k+r})\leq0.80\}.
\end{equation}
Otherwise the trajectory is censored at its last valid observation. This persistence requirement reduces single-observation excursions but uses subsequent measurements to confirm the event retrospectively. Those confirming measurements are not admissible prediction covariates at the earlier inspection. The official MATR EOL convention is different: the first discharge capacity below $0.88$\,Ah, or 80\% of nominal 1.1\,Ah. It yields 83 official events in the 139-cell MATR set, compared with 43 events under pooled harmonisation. The MATR prediction benchmark retains its official endpoint and uses 97 training cells and 42 held-out cells.

Lagged SOH, historical capacity decline, inspection interval and operating information when recorded are candidate covariates. Continuous predictors are standardised on training cells only. Operating descriptors and resistance are not uniformly observed across all sources, so the main pooled effect is reported for lagged SOH; details of temporal availability appear in Supplementary Table~S1. The person-period rows terminate at endpoint or censoring. Separate inspections from the same cell are never divided across training and test folds.

\section{Grouped proportional-hazards formulation}\label{sec:method}
\subsection{Hazard likelihood and coefficient scale}
Let $I_j=(t_{j-1},t_j]$ denote an inspection interval and $y_{ij}$ equal one for the first confirmed EOL interval of cell $i$ and zero for preceding at-risk intervals. Given covariates $\mathbf x_{ij}$ observed before the interval, the conditional event probability is
\begin{align}
 h_{ij}&=\Pr(T_i\in I_j\mid T_i>t_{j-1},\mathbf x_{ij}),\label{eq:haz}\\
 \log\{-\log(1-h_{ij})\}&=a_j+\bm\beta^\top\mathbf x_{ij},\label{eq:cloglog}\\
 h_{ij}&=1-\exp\{-\exp(a_j+\bm\beta^\top\mathbf x_{ij})\}.\label{eq:inv}
\end{align}
Here $a_j$ is the baseline time contribution for the whole inspected interval, not a per-cycle event rate. To see the PH relationship, suppose that covariates are constant within $I_j$ and the instantaneous hazard is $\lambda_i(t)=\lambda_0(t)\exp(\bm\beta^\top\mathbf x_{ij})$. With $\Delta H_{0j}=\int_{t_{j-1}}^{t_j}\lambda_0(u)\,du$, conditional survival through the interval becomes
\begin{equation}\label{eq:grouped_surv}
1-h_{ij}=\exp\{-\Delta H_{0j}\exp(\bm\beta^\top\mathbf x_{ij})\}.
\end{equation}
Consequently $a_j=\log\Delta H_{0j}$ and $\exp(\beta_k)$ is the PH hazard ratio for a one-unit change in covariate $k$ under this model \cite{Cox1972,PrenticeGloeckler1978}. This equality establishes the coefficient scale implied by the model; it is not an empirical test of proportionality over all cycles.

The observed at-risk intervals contribute the Bernoulli log-likelihood
\begin{equation}\label{eq:lik}
\ell(\bm\theta)=\sum_i\sum_{j\in\mathcal R_i}\left[y_{ij}\log h_{ij}+(1-y_{ij})\log(1-h_{ij})\right],\quad
\widehat{\bm\theta}=\arg\max_{\bm\theta}\{\ell(\bm\theta)-\alpha\lVert\bm\beta\rVert_2^2\}.
\end{equation}
The training analysis screened 13 ridge penalties by five-fold whole-cell cross-validation and selected $\alpha=10^{-1.5}$ with the one-standard-error rule. Standardisation makes the reported SOH hazard ratio a per-training-standard-deviation association. A thousand bootstrap samples resampled complete cells and refitted the model. The selected-penalty lagged-SOH estimate was HR = 0.50 (95\% cell-bootstrap CI 0.47--0.52). Higher observed health was associated with lower subsequent interval EOL hazard conditional on the fitted covariates. The estimate is not a causal effect of changing capacity.

\begin{figure}[tbp]\centering
\includegraphics[width=\linewidth]{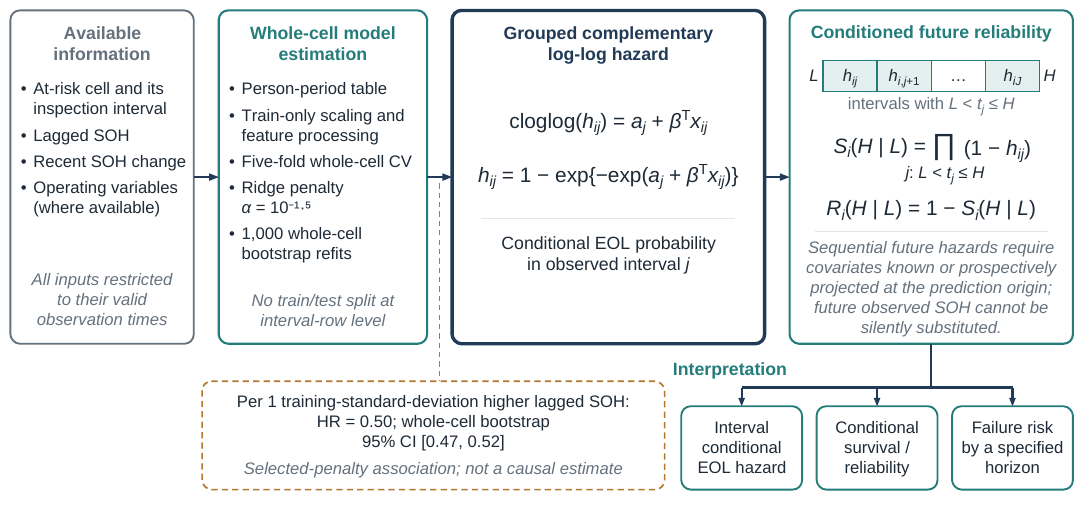}
\caption{Grouped-hazard estimation and conditional reliability calculation. The figure depicts the training-only penalty search, whole-cell resampling and sequential survival calculation; subsequent observed SOH cannot enter a prediction issued at an earlier landmark. The hazard ratio is conditional on the proportional-hazards specification.}\label{fig:dph_workflow}
\end{figure}

For future intervals whose covariates are known or specified prospectively, the conditional reliability and EOL risk from landmark $L$ to horizon $H$ are
\begin{equation}\label{eq:rel}
S_i(H\mid L)=\prod_{j:I_j\subset(L,H]}(1-h_{ij}),\qquad R_i(H\mid L)=1-S_i(H\mid L).
\end{equation}
A reliability curve also defines a discrete residual-life distribution through $\Pr(T_i\in I_j\mid T_i>L)=S_i(t_{j-1}\mid L)h_{ij}$. This construction requires a future hazard path; measured future SOH cannot be inserted after the prediction has been issued. When that path is unavailable, direct landmark regression is a distinct, leakage-free approach to the future-horizon question.

\subsection{Fixed-horizon landmark model and scoring}
The held-out MATR benchmark fits a separate binary cloglog regression at each landmark--horizon pair, not a cumulative-risk curve generated from Eq.~\eqref{eq:rel}. For the indicator $Y_i(L,H)$ of EOL by $H$ among cells still under observation at $L$, it uses only measurements $\mathbf z_i(L)$ available by that landmark:
\begin{equation}\label{eq:landmark}
\log\{-\log[1-\Pr\{Y_i(L,H)=1\}]\}=\gamma_0+\bm\gamma^\top\mathbf z_i(L).
\end{equation}
This direct horizon regression shares the cloglog link and the prospective information-cutoff discipline with grouped hazard modelling, but its coefficients are not the person-period hazard ratios in Eq.~\eqref{eq:cloglog}. The empirical benchmark therefore tests achievable fixed-horizon performance under this related model, not the prediction performance of the accumulated DPH survival function. Landmark prediction follows the dynamic prediction principle of van Houwelingen \cite{VanHouwelingen2007}.

Landmarks are $L\in\{150,200,300\}$ cycles and horizons $H\in\{500,800,1000\}$. The same 42 held-out cells are reused in all nine settings, with known outcomes for 42, 39 and 33 cells at the three respective horizons. A case is scored only if failure is known by $H$ or observation establishes survival past $H$; a cell censored earlier is not coded as a non-event. The Brier score is $n^{-1}\sum_i(y_i-p_i)^2$, log loss is $-n^{-1}\sum_i[y_i\log p_i+(1-y_i)\log(1-p_i)]$, and AUC measures ranking. Censoring-dependent complete-case selection can bias proper scores; the supplementary simulation examines this issue \cite{GerdsSchumacher2006,Heagerty2000}.

\section{Empirical evaluation}\label{sec:results}
\subsection{Leakage-free MATR prediction}
\begin{figure}[tbp]\centering
\includegraphics[width=\linewidth]{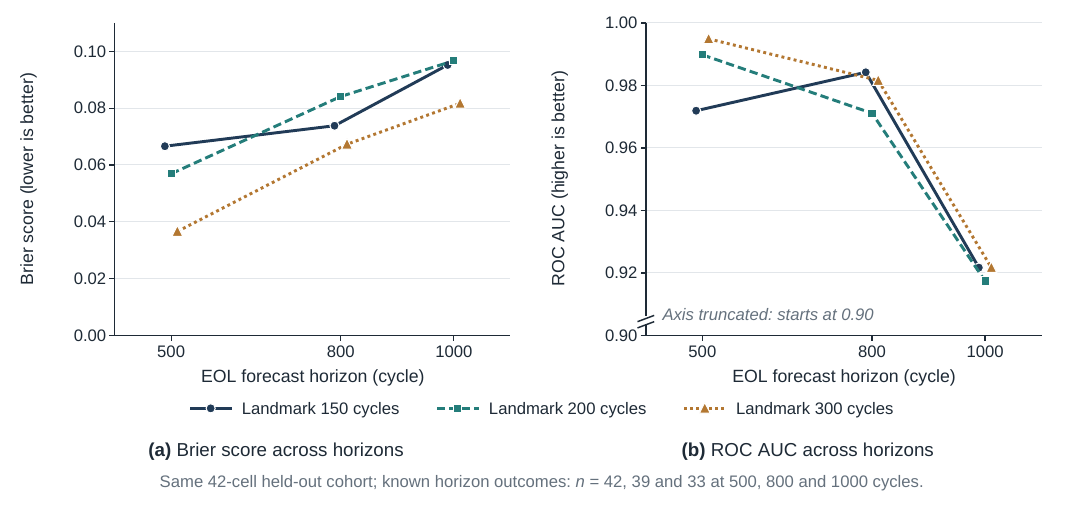}
\caption{Brier score and AUC across fixed landmarks and future horizons in the held-out MATR cohort. The AUC axis is truncated at 0.90 for legibility, whereas the Brier axis starts at zero.}\label{fig:brier}
\end{figure}

Table~\ref{tab:landmark_grid} and Figure~\ref{fig:brier} summarise the nine direct landmark regressions. The best numerical Brier score is 0.0366 at $(L,H)=(300,500)$, while the largest is 0.0967 at $(200,1000)$. These tasks have different lead times, endpoint prevalences and follow-up support. An AUC exceeding 0.97 in several settings shows strong discrimination but not necessarily accurate absolute risk. The full calibration grid is provided in Supplementary Table~S3.
\begin{table}[tbp]\centering\small
\caption{Held-out MATR direct landmark cloglog predictions, using the official $0.88$\,Ah endpoint. All settings reuse one test cohort.}\label{tab:landmark_grid}
\begin{tabular}{rrrrr}\toprule
$L$ & $H$ & Known $n$ & Brier & AUC\\\midrule
150&500&42&0.0666&0.9719\\
150&800&39&0.0738&0.9842\\
150&1000&33&0.0952&0.9217\\
200&500&42&0.0569&0.9898\\
200&800&39&0.0841&0.9711\\
200&1000&33&0.0967&0.9174\\
300&500&42&0.0366&0.9949\\
300&800&39&0.0673&0.9816\\
300&1000&33&0.0817&0.9217\\\bottomrule
\end{tabular}
\end{table}

On the nine matched settings, the random forest attained mean Brier 0.0589, compared with 0.0732 for dynamic-simple landmark cloglog. Table~\ref{tab:models} reports results from that single benchmark comparison. The random forest's better mean probability score prevents a blanket claim of predictive superiority for cloglog. In contrast, the grouped model supplies a conventional coefficient scale and an explicit interval-survival construction. A bootstrap at $(200,500)$ gave mean Brier 0.060 (95\% interval 0.035--0.107) and AUC 0.972 (0.938--0.995); the bootstrap summaries differ from the fixed-split points in Table~\ref{tab:landmark_grid} because they average over resampled cells. These intervals concern the landmark estimator alone, not paired forest--DPH differences.
\begin{table}[tbp]\centering\small
\caption{Mean metrics from the same nine-setting fixed-horizon MATR comparison. Scores are for landmark predictors, not accumulated person-period DPH survival.}\label{tab:models}
\begin{tabular}{lrrrr}\toprule
Predictor & Brier & Log loss & AUC & Mean $|\mathrm{CalErr}|$\\\midrule
Random forest & 0.0589 & 0.1981 & 0.9749 & 0.0219\\
Histogram gradient boosting & 0.0732 & 0.2860 & 0.9731 & 0.0519\\
Dynamic-simple cloglog & 0.0732 & 0.3312 & 0.9616 & 0.0575\\
Multilayer perceptron & 0.1361 & 0.4277 & 0.9598 & 0.0727\\\bottomrule
\end{tabular}
\end{table}

Calibration changes markedly with the decision horizon. At $(150,500)$, mean predicted risk was 0.2331 versus an observed event fraction 0.3333, an absolute difference of 0.1002. At $(200,800)$ the corresponding values were 0.5463 and 0.5128, a difference of 0.0334. By cycle 1000, fitted calibration slopes were between 0.16 and 0.26 across the three landmarks, although only 33 known-outcome cells contribute to each estimate. Decisions that depend on a fixed probability threshold should therefore use horizon-specific calibration evidence rather than AUC alone \cite{VanCalster2019,Bharat2018}.

\subsection{Nonlinear health effects}
An additional matched three-way spline ablation considered a linear SOH term, an unconstrained nonlinear spline and a monotone nonincreasing spline. In its nine landmark settings, mean Brier scores were 0.0951, 0.0858 and 0.0862, respectively; mean log losses were 0.3100, 0.2906 and 0.2892. The monotone fit improved Brier and log loss relative to linear in seven of nine settings and eliminated locally increasing SOH contributions detected in eight of nine unconstrained fits. This separate ablation uses a different refit protocol from Table~\ref{tab:models}; the Brier values are not interchangeable model rows. The main benefit arises from nonlinear representation, while the monotonicity constraint primarily controls the shape of the estimated effect \cite{PyaWood2015}. The paired ablation and its grouped-CV diagnostic are reported in Supplementary Table~S4 and Figure~S3.

\section{Transfer to new sources and chemistries}\label{sec:transfer}
\begin{figure}[tbp]\centering
\includegraphics[width=.82\linewidth]{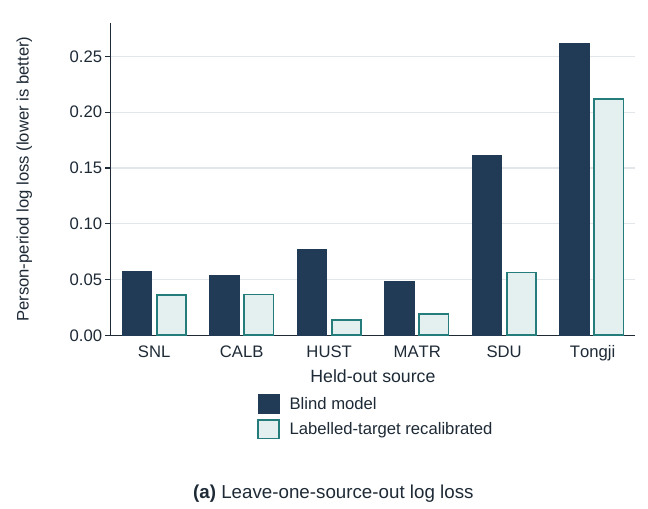}
\caption{Leave-one-source-out person-period log loss before and after labelled-target recalibration. The plotted values refer to interval-event log loss, not a cumulative failure probability by a maintenance horizon.}\label{fig:source_transfer}
\end{figure}

The leave-one-source-out experiment trained on other sources and scored person-period event probabilities on the excluded source. Target-labelled calibration improved interval-level log loss in all six holdouts: on SNL from 0.0573 to 0.0359, on HUST from 0.0769 to 0.0136 and on Tongji from 0.2615 to 0.2117 (Figure~\ref{fig:source_transfer}). These are interval-event scores, which should not be read as cumulative EOL probabilities by a maintenance horizon. Recalibration was fitted on labelled target cells and scored on disjoint target evaluation cells; it is different from blind transfer.

Chemistry holdout exposed a probability-scale discrepancy. The pooled model's mean NMC interval-event probability was 0.0736 against the observed 0.0258; for NCA the corresponding values were 0.1019 and 0.0711 (Table~\ref{tab:chemistry}). A new chemistry can change the baseline risk, covariate distribution or relation between health and EOL. The ability to rank high-risk cells within a source does not ensure correctly scaled failure probabilities in an unseen one.
\begin{table}[tbp]\centering\small
\caption{Blind pooled-model predictions on held-out chemistry, at the person-period level.}\label{tab:chemistry}
\begin{tabular}{lrr}\toprule
Held-out chemistry & Observed interval rate & Mean predicted rate\\\midrule
NMC & 0.0258 & 0.0736\\
NCA & 0.0711 & 0.1019\\
LFP & 0.0109 & 0.0260\\\bottomrule
\end{tabular}
\end{table}

This distinction matters for target recalibration. Repeated chemistry-specific splits using 30\% labelled target cells brought the estimated interval rates close to their empirical frequencies for NMC (0.0268 predicted versus 0.0259 observed) and NCA (0.0732 versus 0.0714). The improvement was not uniform for cell-level fixed-horizon targets. For instance, under SNL source holdout, target recalibration worsened mean fixed-horizon log loss even while improving the person-period score. Effective calibration support depends on the spread and event mixture of target risk scores, not merely the count of labelled cells. A separate information diagnostic compared 24 HUST calibration cells ($n_{\rm eff}\approx0.18$) with 17 NCA cells ($n_{\rm eff}\approx2.18$), indicating markedly different slope information despite the opposite ordering in cell counts. These values refer to those calibration subsets, not to the 30\% chemistry splits reported above. The simulation below investigates slope instability when support is sparse.

\section{Simulation studies}\label{sec:sim}
\subsection{Inspection grouping and link choice}
Event times were generated from a continuous Weibull PH model with known coefficients, inspected at three grid widths and fitted by cloglog, logit and probit. Averaged over 162 factorial settings with 1,000 repetitions, mean coefficient bias was 0.0012 under cloglog, 0.0832 under logit and $-0.2295$ under probit. At the coarsest grid, 95\% coverage of the generating coefficient was 0.946, 0.643 and 0.281, respectively, while horizon-risk errors were all near 0.049 (Table~\ref{tab:link_grid}). Thus comparable probability scores can conceal different coefficient interpretations when the generator is PH.
\begin{figure}[tbp]\centering
\includegraphics[width=\linewidth]{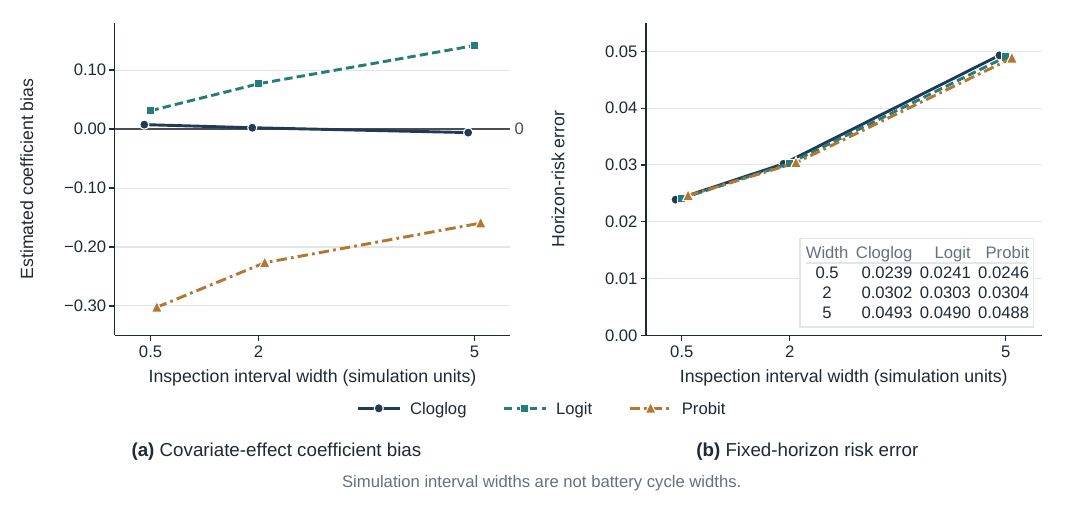}
\caption{Simulated coefficient bias and horizon-risk error by inspected interval width. Widths refer to the simulation time scale, not battery cycles.}\label{fig:link}
\end{figure}
\begin{table}[tbp]\centering\small
\caption{Simulation estimates for cloglog / logit / probit, in that order. Each row reports a common inspection width under the continuous PH generator.}\label{tab:link_grid}
\begin{tabular}{rlll}\toprule
Width & Coefficient bias & 95\% coverage & Horizon-risk error\\\midrule
0.5 & 0.0075 / 0.0311 / $-0.3022$ & 0.9306 / 0.9001 / 0.0029 & 0.0239 / 0.0241 / 0.0246\\
2 & 0.0022 / 0.0768 / $-0.2267$ & 0.9466 / 0.8096 / 0.0714 & 0.0302 / 0.0303 / 0.0304\\
5 & $-0.0060$ / 0.1416 / $-0.1595$ & 0.9460 / 0.6431 / 0.2812 & 0.0493 / 0.0490 / 0.0488\\\bottomrule
\end{tabular}
\end{table}

\subsection{Lead time, censoring and calibration updating}
A separate threshold-proximity simulation fixed the information cutoff one to sixteen intervals before endpoint passage. Mean AUC declined from 0.8878 to 0.8070 and Brier increased from 0.1650 to 0.1903 as the gap widened. A short lead time can therefore make retrospective prediction look better without increasing the information available at an earlier service decision. In a censoring experiment, complete-case Brier bias under observed-covariate-dependent dropout was $-0.0081$; estimated inverse-probability-of-censoring weighting reduced it to $-0.0001$. When dropout depended on unobserved event time, both remained biased. These simulations explain why the later MATR horizon scores, based on fewer observed outcomes, require caution.

For target calibration, let $\eta=\log\{-\log(1-p)\}$ be a blind prediction's cloglog score. Recalibration uses $p^*=1-\exp\{-\exp(a+b\eta)\}$, estimating $a$ and a slope $b$ with quadratic shrinkage $\rho(b-1)^2/2$. Fixing $b=1$ updates the baseline level without fitting a slope, while weaker shrinkage permits a target-specific change in spread \cite{Steyerberg2004}. With 25 labelled target cells under a simulated true slope multiplier of 1.5, mean log loss was 0.4599 for intercept-only, 0.5326 for the free slope and 0.4552 with slope shrinkage; with 50 cells, a free slope performed better (0.4379) than shrinkage (0.4428). These results show a bias--variance trade-off, not a universally optimal numerical cutoff. Full simulation grids, the tested count-based safeguard and the effective-information interpretation are given in the supplement.

\section{Discussion and practical implications}\label{sec:discussion}
The grouped likelihood makes interval-inspected EOL a natural reliability outcome. Under the proportional-hazards specification, the complementary-log-log coefficient retains a cumulative-hazard interpretation; the simulations show that this scale can recover the generating effect even when inspection intervals become wider. Across the six-source battery corpus, higher lagged SOH was associated with lower interval EOL hazard. This hazard ratio summarises how observed battery condition relates to interval EOL hazard \cite{Birkl2017,ZhangSiHuKong2015}.

The fixed-horizon landmark study complements this analysis by assessing what can be predicted using only measurements available at a specified cycle. Its Brier and AUC results describe the landmark cloglog regressions, rather than a sequential forecast derived from the person-period model. Random forest achieved a lower mean Brier score in the matched benchmark, while the grouped cloglog model provides a directly interpretable interval-hazard formulation. These approaches address related but distinct reliability questions. Stochastic degradation models offer another useful representation when the principal target is a residual-life distribution \cite{Han2023,Yang2024,CarrWang2008}.

For practical use, the endpoint convention, available follow-up and target battery population all matter. The later MATR horizon comparisons involve fewer cells with known outcomes, while the source and chemistry holdouts show that absolute-risk calibration can change between populations. The reported hazard ratio is interpreted within the fitted proportional-hazards model. These considerations favour checking calibration for the intended chemistry and decision horizon when extending laboratory-cell findings to new operating conditions.

\section{Conclusion}\label{sec:conclusion}
A grouped complementary-log-log model provides an interpretable representation of capacity-defined EOL observed at discrete inspections. In six-source battery data, higher lagged SOH was associated with lower interval hazard (HR 0.50, 95\% CI 0.47--0.52). Simulations supported the PH coefficient scale under grouping, while independent MATR landmark regressions showed good discrimination but variable calibration and a lower mean Brier score for random forest. Blind chemistry holdout revealed a need for population-specific calibration. Hazard ratios from the grouped model and risks from landmark forecasts answer different questions, and each should be validated at its own observation unit.

\section*{Supplementary material}
Appendix~\ref{app:supp} contains the supplementary material: variable dictionary, interval-history illustration, detailed calibration and spline diagnostics, censoring and recalibration simulation grids, and additional source-transfer tables. Tables, figures and equations numbered with the prefix S are in that appendix.

\section*{Acknowledgements}
The authors thank the researchers and institutions that made the battery ageing datasets used in this study publicly available.

\section*{Author contributions}
Simhaa T. T.: conceptualization, methodology, software, formal analysis, investigation, data curation, visualization, writing (original draft). Sudheesh Kumar Kattumannil: conceptualization, methodology, supervision, validation, writing (review and editing).

\section*{Statements and declarations}
\subsection*{Ethical considerations}\noindent The analysis used publicly available, non-human experimental battery data; ethical approval was not applicable.
\subsection*{Consent to participate}\noindent Not applicable.
\subsection*{Consent for publication}\noindent Not applicable.
\subsection*{Declaration of conflicting interest}\noindent The authors declare no conflicts of interest.
\subsection*{Funding statement}\noindent This research received no specific grant from any funding agency in the public, commercial or not-for-profit sectors.
\subsection*{Data and code availability}\noindent Source data are accessible through the MATR portal (\url{https://data.matr.io/1/}) and the BatteryLife release (\url{https://github.com/Ruifeng-Tan/BatteryLife}). Processed analytical outputs and implementation code will be made available upon acceptance, subject to the source-data terms.

\bibliographystyle{unsrt}
\bibliography{ref}

\clearpage
\appendix
\setcounter{section}{0}\setcounter{table}{0}\setcounter{figure}{0}\setcounter{equation}{0}
\renewcommand{\thesection}{S\arabic{section}}
\renewcommand{\thetable}{S\arabic{table}}
\renewcommand{\thefigure}{S\arabic{figure}}
\renewcommand{\theequation}{S\arabic{equation}}
\section*{Appendix: Supplementary material}\label{app:supp}
\phantomsection

\noindent The main article reports the six-source endpoint audit, full prediction results, calibration, nonlinear SOH comparisons and transfer results. The supplementary figures describe event-history construction, landmark information cutoffs, nonlinear SOH effects and calibration sensitivity. This supplement gives the additional numerical grids needed to audit those analyses. All entries use the documented model variant or simulation design and are not cross-compared when their evaluation units differ.

The underlying battery data were originally published as the MATR and BatteryLife resources \cite{Severson2019,Tan2025BatteryLife}. The main manuscript states which endpoint is evaluated in each analysis.

\section{Variable definition and prospective inspection timeline}
\begin{table}[htbp]\centering\footnotesize
\caption{Variable dictionary and temporal-availability constraints for the grouped and landmark analyses.}\label{stab:variables}
\begin{tabularx}{\linewidth}{p{.26\linewidth}X p{.21\linewidth}}\toprule
Variable & Definition and temporal restriction & Role\\\midrule
Cell identifier & Stable unique cell key across all longitudinal rows & Split and cluster unit\\
Cycle interval & Inspected-cycle span, with start, end and midpoint & Exposure and baseline time\\
At-risk status & At risk before the interval and not already censored & Likelihood inclusion\\
Interval event & First harmonised endpoint in interval, otherwise zero & Grouped-hazard outcome\\
Lagged SOH & Capacity fraction evaluated before the scored interval & Primary interpretable health effect\\
Recent SOH change & Historical difference over available preceding cycles & Past-only degradation dynamics\\
Operating conditions & Temperature, charging rate and other protocol descriptors when available & Potential covariates, not pooled causal effects\\
Resistance & Previous DC-resistance reading where recorded & Optional health indicator\\
Source/chemistry & Dataset provenance and electrode classification & Transfer strata, partial confounding\\
Landmark state & Measurements no later than cycle $L$ & Fixed-horizon prediction\\\bottomrule
\end{tabularx}
\end{table}
\begin{figure}[H]\centering
\includegraphics[width=\linewidth]{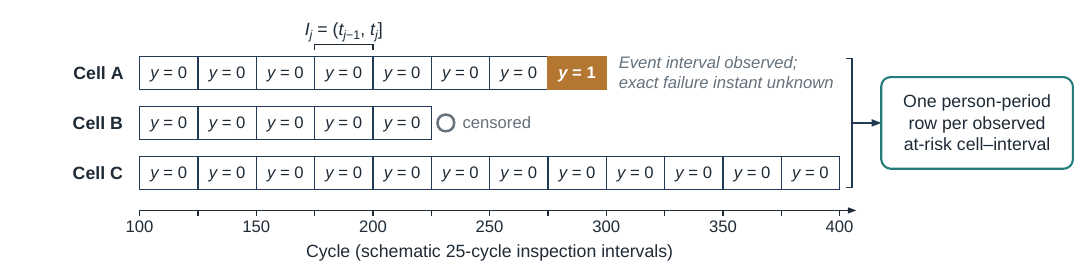}
\caption{Schematic person-period expansion of inspected battery cell histories. A cell leaves the risk set after endpoint or censoring; no post-event observations enter estimation.}\label{sfig:intervals}
\end{figure}
\begin{figure}[H]\centering
\includegraphics[width=\linewidth]{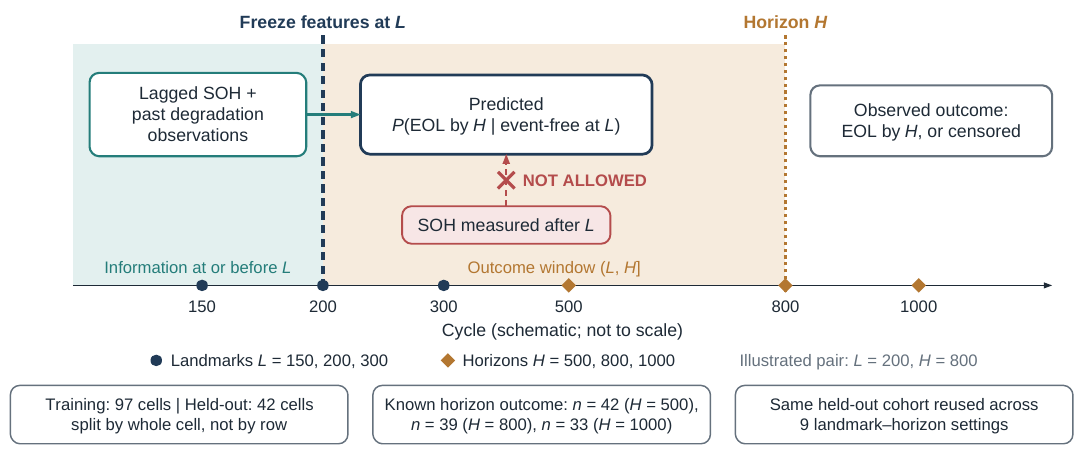}
\caption{Permitted information at landmark $L$ and subsequent outcome evaluation at horizon $H$. The shaded future observation region is never used to construct landmark features.}\label{sfig:landmark_timeline}
\end{figure}
The pooled hazard analysis used nominal 25-cycle intervals after cycle 100. Endpoint and grouping sensitivities examined 10-, 25- and 50-cycle grids; the 25-cycle representation was retained to balance inspection resolution and event support. Numeric coefficient intervals in this supplement refer to explicitly stated fitted variants only.

\section{Held-out cell censoring audit}
\begin{table}[htbp]\centering\scriptsize
\caption{Cell-level horizon-censoring audit for the nine MATR landmark--horizon predictions. Censoring counts are relative to the 42-cell cohort at each fixed horizon.}\label{stab:censor_audit}
\begin{tabular}{rrrrrrr}\toprule
$L$ & $H$ & At risk at $L$ & Events by $H$ & Known survivors & Censored before $H$ & Used \\\midrule
150 & 500 & 42 & 14 & 28 & 0 & 42\\
150 & 800 & 42 & 20 & 19 & 3 & 39\\
150 & 1000 & 42 & 23 & 10 & 9 & 33\\
200 & 500 & 42 & 14 & 28 & 0 & 42\\
200 & 800 & 42 & 20 & 19 & 3 & 39\\
200 & 1000 & 42 & 23 & 10 & 9 & 33\\
300 & 500 & 42 & 14 & 28 & 0 & 42\\
300 & 800 & 42 & 20 & 19 & 3 & 39\\
300 & 1000 & 42 & 23 & 10 & 9 & 33\\\bottomrule
\end{tabular}
\end{table}

These counts use the same 42 held-out MATR cells at all prediction landmarks. At horizon 500 every selected cell has a known event status; three and nine cells, respectively, lack a known binary status by 800 and 1000 cycles. Exclusion of those cells changes the estimand of the reported complete-case scores. The primary article reports the nine Brier scores, AUCs and calibration diagnostics.

\section{Extended empirical calibration and SOH-shape results}
\begin{table}[htbp]\centering\scriptsize
\caption{Landmark calibration diagnostics for the nine held-out MATR settings. All rows reuse the same cohort.}\label{stab:calibration_grid}
\begin{tabular}{rrrrrrrr}\toprule
$L$ & $H$ & $n$ & Events & Mean risk & Observed rate & Absolute error & Cal. slope\\\midrule
150&500&42&14&0.2331&0.3333&0.1002&1.1876\\
150&800&39&20&0.5518&0.5128&0.0390&2.1461\\
150&1000&33&23&0.7401&0.6970&0.0431&0.2510\\
200&500&42&14&0.2468&0.3333&0.0866&1.3343\\
200&800&39&20&0.5463&0.5128&0.0334&1.1178\\
200&1000&33&23&0.7528&0.6970&0.0558&0.1717\\
300&500&42&14&0.2886&0.3333&0.0447&3.1080\\
300&800&39&20&0.5796&0.5128&0.0668&0.3482\\
300&1000&33&23&0.7452&0.6970&0.0482&0.1635\\\bottomrule
\end{tabular}
\end{table}
Across the nine fitted settings, the observed horizon-event fractions were 0.3333, 0.5128 and 0.6970 at cycles 500, 800 and 1000, respectively. These diagnostics estimate the calibration of a fixed-horizon landmark regression rather than the accumulated person-period survival curve.
\begin{table}[htbp]\centering\small
\caption{Separate matched SOH-shape ablation at the selected penalty; its fitted variants differ from the main benchmark comparator set.}\label{stab:spline_full}
\begin{tabular}{llrrrr}\toprule
Design & Model & Brier & Log loss & AUC & Increasing fits\\\midrule
\multirow{3}{*}{Landmark, 9 settings} & Linear & 0.095050 & 0.309975 & 0.969664 & n/a\\
 & Unconstrained & 0.085794 & 0.290630 & 0.964666 & 8/9\\
 & Monotone & 0.086198 & 0.289172 & 0.972563 & 0/9\\\midrule
\multirow{3}{*}{Grouped CV, 5 folds} & Linear & 0.017958 & 0.090692 & 0.763031 & n/a\\
 & Unconstrained & 0.015646 & 0.060861 & 0.984220 & 5/5\\
 & Monotone & 0.016398 & 0.066831 & 0.985842 & 0/5\\\bottomrule
\end{tabular}
\end{table}
The shape ablation comprises separate matched models fitted with linear, unconstrained spline and monotone spline SOH terms. It is not interchangeable with the broader random-forest/gradient-boosting comparator benchmark in the main article.

\section{Shape-constraint simulation and penalty sensitivity}

Under a smoothly varying monotone generating function (216 factorial cells, generally 1,000 replicates), the unconstrained spline had the lowest integrated mean squared error (IMSE) and risk mean absolute error (MAE); a constraint was not free of estimation cost (Table~\ref{stab:shape}). With deliberately mildly non-monotone truth, unconstrained IMSE was about 0.589 versus 0.687 for the monotone estimator. Thus a monotone physical assumption must itself be justified.
\begin{table}[htbp]\centering\small
\caption{Shape-constraint simulation: aggregate errors under monotone truth with ordinary penalisation.}\label{stab:shape}
\begin{tabular}{lrrr}\toprule
Estimator & IMSE & Risk MAE & Increasing-fit frequency\\\midrule
Linear & 0.722741 & 0.142019 & Not applicable\\
Unconstrained spline & 0.443630 & 0.116108 & 0\\
Monotone spline & 0.527347 & 0.130966 & 0 (enforced)\\\bottomrule
\end{tabular}
\end{table}

\begin{figure}[!tbp]
\centering
\includegraphics[width=\linewidth]{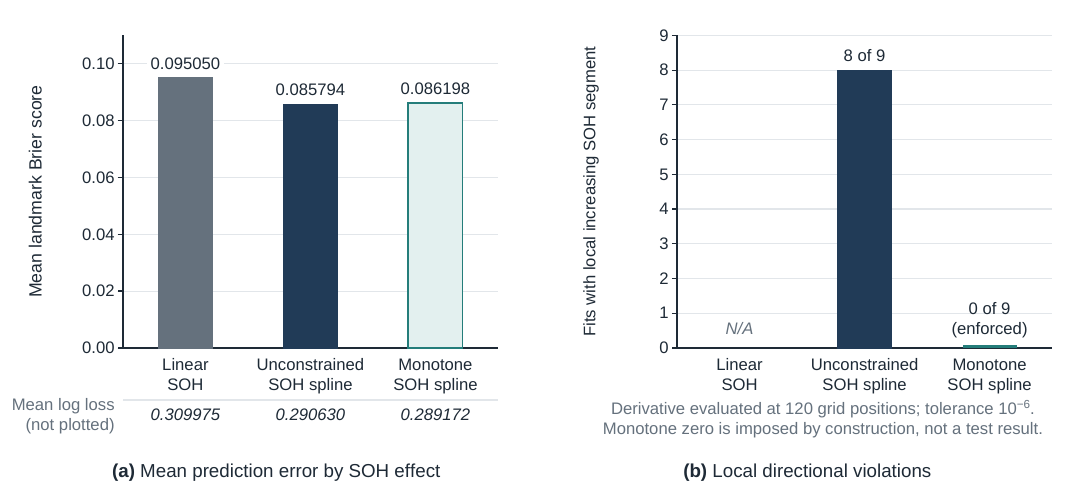}
\caption{Observed mean landmark prediction error for linear, unconstrained-spline and monotone-spline SOH effects (a), and the number of fitted unconstrained curves with locally increasing sections compared with the constrained curves (b). The bar comparison is not a test of statistical separation between the two spline fits.}
\label{sfig:soh_comparison}
\end{figure}

A second 32-cell by 500-replicate sweep exposed a conditional value. At a very small penalty ($10^{-6}$), 100 cells and 20\% censoring, monotone IMSE was 0.1610 against 0.7128 for unconstrained, with 35.2\% unconstrained fits increasing over some portion of the curve. Under 50\% censoring, the corresponding IMSEs were 0.2013 against 0.8064 with 41.8\% reversals. Once $\alpha\geq10^{-4}$ the primary smooth simulation did not show the same systematic advantage, in contrast to observed heterogeneous SOH trajectories.

The shape simulation uses a smooth analytic truth generated outside the spline basis. On its principal grid the unconstrained spline had lower IMSE than the monotone fit (0.443630 versus 0.527347) and lower risk MAE (0.116108 versus 0.130966); both improved over the linear fit (IMSE 0.722741). Under the chosen ridge penalty the unconstrained simulations did not develop derivative reversals, so the monotone constraint had nothing to correct in that particular generating mechanism. With mildly non-monotone truth the constraint cost grew (IMSE approximately 0.687 versus 0.589). The real corpus is less smooth and more heterogeneous, which explains why reversals appear there. The weak-regularisation follow-up tests when a constraint becomes useful in simulation rather than obscuring the negative primary result.

\begin{table}[htbp]\centering\small
\caption{Selected paired weak-regularisation results from Study A follow-up. Lower IMSE is better; violation share is the fraction of unconstrained fits with an increasing segment.}\label{stab:weakspline}
\begin{tabular}{rrrrrr}\toprule
Censoring & Cells & Penalty & Unconstr. IMSE & Monotone IMSE & Violation share\\\midrule
20\%&100&$10^{-6}$&0.7128&0.1610&0.352\\
50\%&100&$10^{-6}$&0.8064&0.2013&0.418\\
20\%&500&$10^{-6}$&0.1512&0.0710&0.122\\
50\%&500&$10^{-6}$&0.1636&0.0719&0.132\\
20\%&100&$10^{-4}$&0.1303&0.1957&0.026\\
50\%&500&0.003162&0.0026&0.0286&0.000\\
20\%&100&0.031623&0.3577&0.4656&0.000\\\bottomrule
\end{tabular}
\end{table}
\noindent Under nearly unpenalised fits, monotonicity reduces IMSE sharply and suppresses reversals. At conventional positive ridge penalties the same constraint can add bias when the unconstrained curve is already stable. This conditional result explains the otherwise conflicting evidence between the smooth simulation and the heterogeneous real data.

\section{Lead-time and censoring simulations}
\begin{table}[htbp]\centering\small
\caption{Event-proximity simulation, 135 configurations with 1,000 replicates each. Gap is the number of intervals ahead of the endpoint where prediction is issued.}\label{stab:gap}
\begin{tabular}{rrrr}\toprule
Gap (intervals) & AUC & Brier & Log loss\\\midrule
1 & 0.8878 & 0.1650 & 0.5068\\
2 & 0.8862 & 0.1657 & 0.5081\\
4 & 0.8813 & 0.1680 & 0.5128\\
8 & 0.8664 & 0.1746 & 0.5276\\
16 & 0.8070 & 0.1903 & 0.5634\\\bottomrule
\end{tabular}
\end{table}
The study shows that late landmark information inflates apparent accuracy relative to an earlier maintenance decision. This motivates freezing all features at the chosen landmark.
\begin{table}[htbp]\centering\small
\caption{Evaluation bias under censoring, 54 factorial configurations with 1,000 replicates. Values are complete-case / inverse-probability-of-censoring-weighted (IPCW) bias.}\label{stab:ipcw}
\begin{tabular}{lrrr}\toprule
Mechanism & Brier bias & Log-loss bias & AUC bias\\\midrule
Independent & 0 / 0 & 0 / 0 & $-0.0000$ / 0.0001\\
Covariate-dependent & $-0.0081$ / $-0.0001$ & $-0.0194$ / $-0.0001$ & $-0.0126$ / $-0.0003$\\
Latent-event-dependent & $-0.0252$ / $-0.0206$ & $-0.0606$ / $-0.0495$ & $-0.0001$ / 0.0032\\\bottomrule
\end{tabular}
\end{table}
For covariate-dependent censoring, the fitted censoring model substantially reduced bias. Larger residual bias remained when censoring depended on unobserved event time. The MATR landmark results are therefore presented for the subsets with known horizon outcomes, without attributing their censoring mechanism to any particular simulation scenario.

The proximity simulation varied the distance between the information cutoff and the threshold event, holding the simulated data-generating mechanism observable for a complete audit. Its grid crossed five event gaps with three levels each of health-signal strength, measurement noise and inspection width. Mean AUC fell from 0.8878 at one interval of lead to 0.8070 at sixteen; Brier rose from 0.1650 to 0.1903 and log loss from 0.5068 to 0.5634. This demonstrates why a near-threshold reading can yield apparently excellent retrospective hazard classification without providing equally useful early prognosis. The battery analysis prevents this shortcut by using fixed landmarks, but its resulting scores still depend on the chosen lead time.

The censoring simulation first generated the full outcome, then removed follow-up under independent, covariate-dependent and latent-event-dependent censoring mechanisms. Estimated inverse-probability-of-censoring weights were obtained from a fitted censoring model, not oracle probabilities. Under covariate-dependent censoring, complete-case Brier bias was $-0.0081$ and weighted bias was $-0.0001$; log-loss bias changed from $-0.0194$ to $-0.0001$. Under latent event-time censoring, Brier bias remained $-0.0206$ even after weighting. This distinction is important: weighting can correct a bias only to the extent that observed variables explain the censoring process. The empirical MATR analysis consequently retains its clearly identified evaluable cohorts rather than substituting simulation-derived corrections.

\section{Additional transfer and recalibration sensitivity}
\begin{table}[htbp]\centering\small
\caption{Illustrative calibration-slope information audit on two distinct labelled-target subsets. $n_{\rm eff}$ characterises information for slope updating; it is not a count of target cells.}\label{stab:neff}
\begin{tabular}{lrr}\toprule
Calibration subset & Labelled target cells & $n_{\rm eff}$\\\midrule
HUST & 24 & 0.18\\
NCA & 17 & 2.18\\\bottomrule
\end{tabular}
\end{table}
The subsets in Table~\ref{stab:neff} are different from the 30\% target-chemistry calibration partitions. The larger HUST cell count carried less slope information, illustrating why an operational calibration update should inspect the score distribution and endpoint mix rather than rely on cell count alone.

In the calibration experiment, the target population had baseline shift $\Delta\in\{0,0.5,1,2\}$ and slope multiplier $\lambda\in\{0.5,1,1.5\}$. Target sets of 50 or 200 cells were split into labelled calibration subsets of size 3, 10, 25 or 50, and held-out scoring subsets. A calibration \emph{fraction} cannot summarise that support: 10\% of 50 cells and 10\% of 200 cells contain very different information. One-class calibration samples were counted as non-identifiable rather than being silently dropped from the denominator. Across 720 factorial configurations, unrestricted two-parameter recalibration had mean log loss 3.4299 at three calibration cells, versus 0.6086 for intercept-only and 0.6231 for blind use. Even at 25 cells the unrestricted mean loss (0.5532) remained above intercept-only (0.4807) when averaging over slope-shift regimes. The $\lambda$-specific breakdown identifies the important exception: if the true slope differs substantially and support is large enough, slope correction can genuinely help.

\begin{table}[htbp]\centering\scriptsize
\caption{Study B recalibration by target support, averaged across the full shift grid (500 replicates per design).}\label{stab:b_summary}
\begin{tabular}{rrrrrr}\toprule
Calibration cells & Blind LL & Intercept LL & Free-slope LL & Free-slope valid share & Free-slope SD\\\midrule
3 & 0.6231 & 0.6086 & 3.4299 & 0.5387 & 14.2666\\
10 & 0.6230 & 0.5017 & 1.0248 & 0.8927 & 6.3004\\
25 & 0.6236 & 0.4807 & 0.5532 & 0.9876 & 2.1104\\
50 & 0.6209 & 0.4687 & 0.4726 & 0.9997 & 0.8176\\\bottomrule
\end{tabular}
\end{table}

\subsection{Support-adaptive slope shrinkage across the simulation grid}
The simulated intercept-only, unrestricted and shrunk-slope estimators were compared at true slope scaling $\lambda=1.5$. Table~\ref{stab:shrunk_full} gives the full sensitivity grid at three scaling factors. This is a simulation of labelled-target update rules, not a prospective trial with incoming cells.
\begin{table}[htbp]\centering\scriptsize
\caption{Support-adaptive slope shrinkage (500 repetitions per design). The shrinkage procedure coincides with intercept-only at 3 and 10 calibration cells.}\label{stab:shrunk_full}
\begin{tabular}{rrrrrr}\toprule
True slope $\lambda$ & Cells & Intercept LL & Free-slope LL & Shrunk LL & Shrunk slope SD\\\midrule
0.5&3&0.6972&4.3137&0.6972&0.00\\
0.5&10&0.5530&1.0479&0.5530&0.00\\
0.5&25&0.5106&0.5631&0.5108&0.20\\
0.5&50&0.4993&0.5071&0.4991&0.27\\
1.0&3&0.6173&3.4341&0.6173&0.00\\
1.0&10&0.5065&1.0936&0.5065&0.00\\
1.0&25&0.4758&0.5631&0.4765&0.16\\
1.0&50&0.4621&0.4772&0.4627&0.24\\
1.5&3&0.5560&2.7309&0.5560&0.00\\
1.5&10&0.4840&1.0670&0.4840&0.00\\
1.5&25&0.4599&0.5326&0.4552&0.27\\
1.5&50&0.4492&0.4379&0.4428&0.50\\\bottomrule
\end{tabular}
\end{table}

\begin{figure}[!tbp]
\centering
\includegraphics[width=\linewidth]{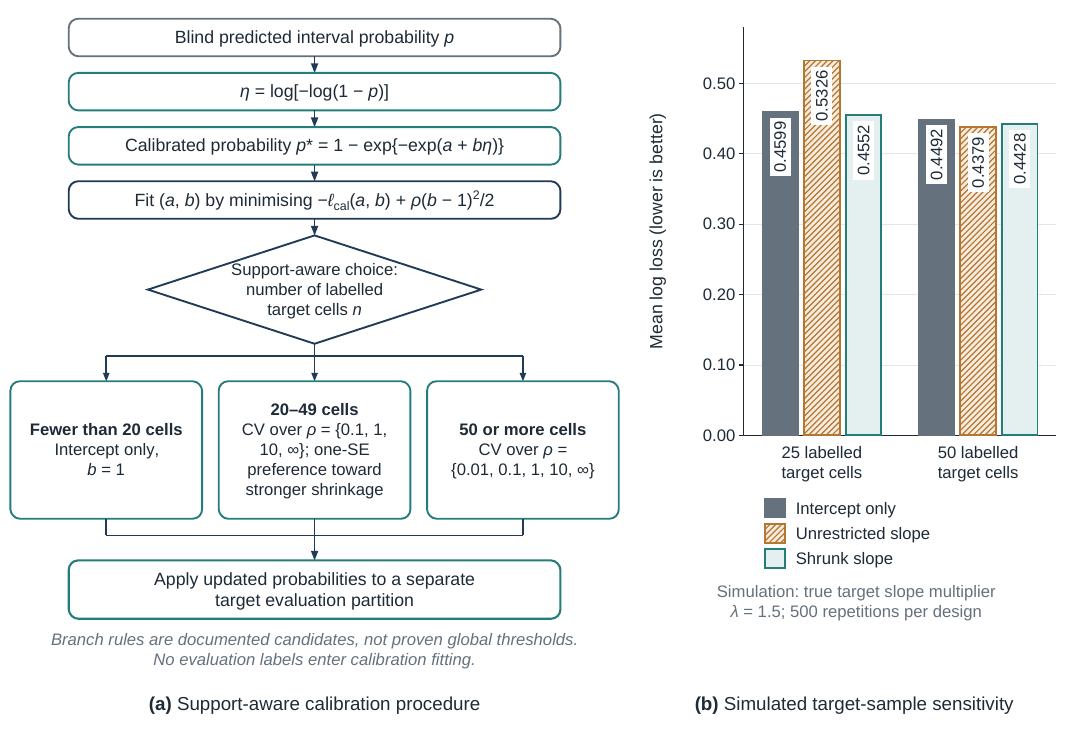}
\caption{Support-adaptive calibration procedure (a) and simulation log-loss comparison for 25 and 50 labelled target cells under true slope factor 1.5 (b). The comparison uses 500 simulation repetitions per design and does not constitute prospective deployment evidence.}
\label{sfig:adaptive_recalibration}
\end{figure}

The support-adaptive objective is $-\ell_{\rm cal}(a,b)+\rho(b-1)^2/2$. The limit $\rho=\infty$ fixes $b=1$; small positive $\rho$ allows target slope correction. For fewer than 20 labelled cells, the procedure sets $\rho=\infty$. For 20--49 it cross-validates $\{0.1,1,10,\infty\}$; at 50 or more it also admits $0.01$. Stronger shrinkage is preferred by the one-standard-error rule when calibration-set log losses are indistinguishable. Exactly zero is excluded from the tuning grid. The shrinkage study completed 900 design cells with 500 repeats per cell and no reported fit failures.

\begin{figure}[!tbp]
\centering
\includegraphics[width=\linewidth]{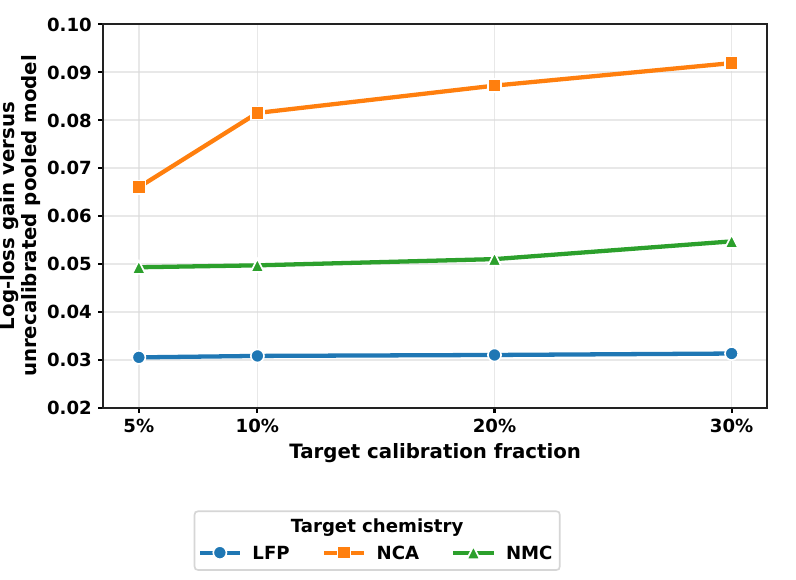}
\caption{Empirical labelled-target calibration-fraction sensitivity. Reductions in person-period log loss are shown for NMC, NCA and LFP target-cell fractions; this figure is distinct from the support-adaptive simulation and does not establish prospective fixed-horizon calibration.}
\label{sfig:legacy_cal_frac}
\end{figure}

The 5--30\% labelled-target fraction sweep is shown in Figure~\ref{sfig:legacy_cal_frac} for visual comparison. It is an empirical person-period assessment and does not validate the numerical support thresholds of the subsequent simulation.
\section{Data sources and reproducibility}
The external-source records are documented by Severson et al.~\cite{Severson2019} and the BatteryLife repository~\cite{Tan2025BatteryLife}. Analyses use whole-cell fold assignment, train-only standardisation, the stated ridge penalty search, and cell-level resampling. The underlying source datasets are public, and processed analysis outputs and implementation code are intended for release on acceptance subject to the source-data licences.

\section{Implementation and computational sensitivity}
\begin{table}[htbp]\centering\small
\caption{Timing and memory measurements from ten runs at 493 cells. Hardware-specific observations, not a universal complexity claim.}\label{stab:runtime}
\begin{tabular}{lrrrr}\toprule
Estimator & Fit (s) & Infer (s) & Recalibrate (s) & Peak memory (MB)\\\midrule
Cloglog & 0.0444 & 0.0002 & 0.0063 & 1.1703\\
Histogram gradient boosting & 0.4399 & 0.0171 & 0.0073 & 3.4413\\\bottomrule
\end{tabular}
\end{table}
These timing and peak-memory observations come from ten implementations at a 493-cell size and are not measures of an algorithm's asymptotic complexity. They are presented only to document the tested implementation.

\end{document}